\documentclass[notitlepage,superscriptaddress,showpacs,nobalancelastpage,twocolumn,aps,longbibliography,prl]{revtex4-2}
\usepackage[english]{babel}
\RequirePackage[T1]{fontenc}
\RequirePackage{times} % favourite for note.

\usepackage[italicdiff]{physics}
\usepackage{siunitx} % typesets numbers with units very nicely
\usepackage{amsfonts}
\usepackage{lipsum}
\usepackage{amsmath}
\usepackage{amssymb}
\usepackage{wasysym}
\usepackage{xspace}
\usepackage{array}
\usepackage[hidelinks]{hyperref}
\usepackage[dvipsnames]{xcolor}
\usepackage{my_symbols}
\usepackage{CJK}
\usepackage{bm}
\usepackage{verbatim}
\usepackage{tikz}
\usepackage{marginnote}
\usepackage[textwidth=1.5cm]{todonotes}
\usepackage[normalem]{ulem}
\usepackage{soul}
\usepackage[commandnameprefix=ifneeded]{changes}

\sethlcolor{green}

\newcommand{\chiB}{\ensuremath{\chi_\ssf{B}}}
\newcommand{\chiT}{\ensuremath{\chi_\ssf{T}}}

{\par}

\begin{document}
\begin{CJK*}{UTF8}{gbsn} % Use default fonts from CJK (see below)
\title{Application of Probabilistic-bit in Precision Measurements}
\author{Yunwen Liu (刘允文)}
\affiliation{Department of Physics and State Key Laboratory of Surface Physics, Fudan University, Shanghai 200433, China}
\author{Jiang Xiao (萧江)}
\email[Corresponding author:~]{xiaojiang@fudan.edu.cn}
\affiliation{Department of Physics and State Key Laboratory of Surface Physics, Fudan University, Shanghai 200433, China}
\affiliation{Institute for Nanoelectronic Devices and Quantum Computing, Fudan University, Shanghai 200433, China}
\affiliation{Shanghai Research Center for Quantum Sciences, Shanghai 201315, China}
\affiliation{Shanghai Branch, Hefei National Laboratory, Shanghai 201315, China}

\begin{abstract}

We propose a novel approach for precision measurement utilizing an ensemble of probabilistic bits (p-bits). This method leverages the inherent fluctuations of p-bits to achieve high sensitivity in various applications, including magnetic field sensing, temperature monitoring and timekeeping. The sensitivity scales proportionally to the square root of the total number of p-bits, enabling unprecedented accuracy with large ensembles. Furthermore, the robustness of this method against device imperfections and non-uniformity enhances its practicality and scalability. This work paves the way for a new paradigm in precision measurement, offering a cost-effective and versatile alternative to traditional methods.

\end{abstract}

\maketitle
\end{CJK*}

\emph{Introduction -}
The scientific discovery and technological innovation are fundamentally anchored in our ability to measure the universe around us. Classical measurement, with its deterministic framework rooted in the laws of classical physics, has long served as the bedrock of this quest, enabling the precise quantification of physical quantities like length, mass, velocity, and temperature. This approach has facilitated a direct and predictable relationship between measurement and physical reality, allowing scientists and engineers to describe, predict, and manipulate the macroscopic world with unparalleled accuracy. The deterministic nature of classical measurement has not only profoundly shaped our understanding of the universe but has also laid the groundwork for the technologies that define modern society.

However, as we venture further into the microscopic and quantum realms, the classical framework begins to reveal its limitations. This is where quantum computation \cite{nielsen_quantum_2010} and quantum precision measurement \cite{degen_quantum_2017}, leveraging the unique properties of qubits, herald a new era of exploration and understanding. Qubits, the fundamental units of quantum information, exhibit extraordinary characteristics such as superposition, entanglement, and squeezing. These properties enable the execution of complex calculations and the attainment of measurement precision on the quantum scale, far beyond the reach of classical techniques.

Quantum precision measurement \cite{degen_quantum_2017}, in particular, stands as a testament to the groundbreaking progress in our capability to probe the natural world. By exploiting the peculiar principles of quantum mechanics, such as entanglement and squeezing \cite{giovannetti_quantum-enhanced_2004}, it is possible to detect and quantify physical phenomena with an accuracy that transcends the classical limits. This enhanced sensitivity, achievable through the meticulous preparation and manipulation of quantum states, has practical implications across a spectrum of fields, including metrology \cite{giovannetti_quantum_2006}, navigation \cite{feng_navigation_2019}, timing \cite{riehle_atom_2017, beattie_cs_2023}, and the detection of gravitational waves \cite{ligo_scientific_collaboration_and_virgo_collaboration_observation_2016}. 

In parallel with these quantum advancements, probabilistic computing emerges as a paradigm shift \cite{camsari_dialogue_2021}. This innovative computing model is underpinned by the probabilistic bit \cite{kaiser_probabilistic_2021}, or p-bit, which embodies a fluctuating state governed by a certain probability distribution. Unlike the binary bit of classical computing, which exists in a definite state of 0 or 1, or the qubit of quantum computing, which leverages phase coherence, the p-bit captures the probabilistic nature of the real world without the complexity of quantum mechanics. This unique characteristic enables p-bits to encode and process information in a manner that mirrors the inherent fluctuations of physical systems, thereby facilitating the efficient simulation of complex systems, optimization problems, and stochastic processes \cite{borders_integer_2019,camsari_p-bits_2019, chowdhury_full-stack_2023}.

In this Letter, we endeavor to broaden the scope of p-bits by exploring their potential in the domain of measurements. This exploration is predicated on the innovative concept of harnessing an ensemble of p-bits for probabilistic precision measurement, a method that stands to significantly enhance the accuracy of measurements under conventional conditions, such as room temperature. 
The practicality of deploying p-bits in the precise detection and measurement of nuanced phenomena, such as the detection of minute magnetic fields, the monitoring of subtle temperature variations, and the provision of robust mechanisms for timekeeping, is thoroughly examined. 

\emph{Principle of Probabilistic Measurement -}
A physical implementation of a probabilistic-bit (p-bit) can be viewed as a double well system \cite{chowdhury_full-stack_2023,kim_biristor_2024,chou_Ising_2019}, where the two stable states (encoded as 0 or 1) are separated by a potential barrier, whose height $\Delta$ is comparable to the thermal activation energy $k_BT$: $\Delta/k_BT \lesssim 10$, such that the physical state undergoes an everlasting fluctuation between the two states. The relative probability of residing in the 0 or 1 state depends on the energy bias $2U$ between the two wells according to the Boltzmann distribution:
\[ \frac{p_1}{p_0} = e^{-2\beta U}
\qor 
p_0 - p_1 = \tanh(\beta U), \]
in which $\beta=1/k_BT$. Consider an ensemble of $N$ independent p-bits, and the population imbalance between 0s and 1s is expected to be
\begin{equation}
    \label{eqn:N0N1}
    n \equiv N_0 - N_1 = N\tanh(\beta U),
\end{equation}
which grows linearly as the function of p-bit number $N$. By measuring this number of p-bit imbalance, it is possible to extract information about $\beta$ or $U$. Since the p-bits keep fluctuating, the measured number $n$ also fluctuates. However, the uncertainty in $n$ due to this fluctuation is proportional to the square root of the total number: $\delta n \propto \sqrt{N}$. Therefore, for sufficiently large $N$, the mean value of the population imbalance can always overshadow the uncertainty: $n > \delta n$, thus enabling the p-bit ensemble for the purpose of measurement. 

\begin{figure}[t]
\centering
\includegraphics[width=\columnwidth]{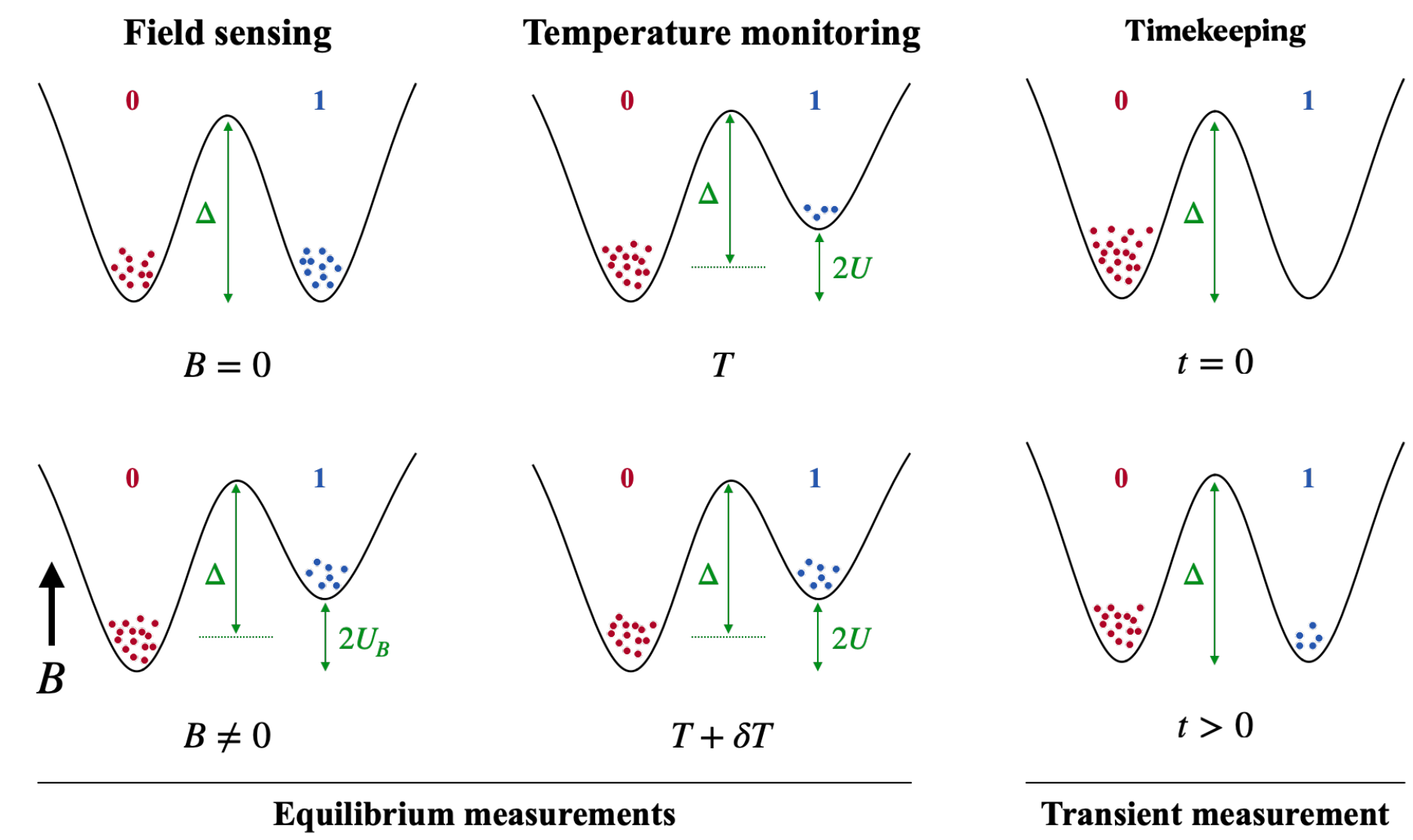}
\caption{Probabilistic bit measurements. Top: reference distributions of the ensemble in zero field $B=0$, initial temperature $T$ and the beginning of timekeeping $t=0$. Bottom: deviated distributions in the presence of magnetic field $B\ne0$, temperature change $\delta T$ and time elapse $t>0$.}
\label{fig:p-measurements}
\end{figure}

Based on this simple principle, we explore the possibility and practicality of probabilistic measurement (p-measurement), and propose strategies that leverage p-bits for various purposes. Compared to some previous works that use noisy or chaotic systems for magnetic field sensing \cite{Grigorenko_sense_1997, silva_chaos_2021}, temperature monitoring \cite{benz_jn_2024} and timekeeping \cite{aitken_dating_1999, lazzaro_c14_2020, mccarthy_time_2018}, the outlined measurement methodologies put emphasis on bistable systems (bits) while preserving the exploitation of randomness. The principles are delineated in \Figure{fig:p-measurements}, where the field sensing and temperature variation monitoring make use of equilibrium states of the p-bit ensemble and the time-keeping relies on transient states. 

\emph{Field sensing -}
For magnetic field sensing, an ensemble of p-bits initially assume a state of equilibrium characterized by an even distribution of 0s and 1s (with $U = 0$), albeit within a certain statistical margin of error. Upon the introduction of an external magnetic field denoted as $B$, an energy bias $U = 2MB$ emerges between the state-0 and the state-1, prompting an asymmetry in the quantities $N_0$ and $N_1$ as expressed in \Eq{eqn:N0N1}. By measuring the imbalance $n = N_0 - N_1$, it becomes feasible to deduce the intensity of the magnetic field. 

\emph{Temperature monitoring -}
In a manner analogous to field sensing, it is possible to determine the temperature based on the energy bias $U \neq 0$ and the measured quantities $N_0$ and $N_1$. However, this approach necessitates prior knowledge of the energy bias $U$, which is a parameter that may not be readily available and could vary from one p-bit to another. Consequently, it is not practical to directly ascertain the absolute temperature using p-bits. However, it is feasible to accurately gauge the variations in temperature without precise knowledge of potentially non-constant device parameters like $U$. For a collection of p-bits with an energy bias of $U$, the population imbalance $n$ is contingent upon temperature. As the temperature rises, the population in state-1 also increases. By measuring the alteration in population imbalance $\delta n = n_{T'} - n_{T}$, it is possible to infer the temperature change $\delta T = T' - T$. 

\emph{Timekeeping -} 
Timekeeping is yet another task that can be achieved probabilistically. Most timekeeping devices rely on some kind of oscillating phenomena with fixed periods. From the rhythmic swings of a pendulum to the consistent vibrations of quartz crystals and the intricate quantum transitions between energy levels, these mechanisms serve as the foundation for tracking time by counting cycles. 
However, a unique form of timekeeping makes use of the radioactive decay of certain atoms, notably Carbon-14 ($^{14}$C). While this method may not boast the same level of precision as its counterparts, it stands out for its simplicity and resilience against various external disturbances. 
An ensemble of p-bits can emulate the behavior of radioactive atoms. The timekeeping is initialized by resetting all p-bits into one single state (say the 0 state). Afterwards, the ensemble undergoes thermal fluctuation and would eventually approach the thermal equilibrium on a time scale comparable to the thermalization time. Therefore, it is possible to estimate the elapsed time by counting the number of p-bits in state-1. Here the thermalization of p-bit is equivalent to the radioactive decay of $^{14}$C. 
The thermalization time for a p-bit, equivalent to the half-life of $^{14}$C, is dictated by the energy barrier $\tau \propto \exp(\beta \Delta)$ and can vary significantly, ranging from sub-microseconds to years. 

\emph{Realization with MTJ p-bits -}
We now consider a possible realization of the probabilistic measurement utilizing the p-bits based on the magnetic tunnel junction (MTJ) \cite{yan_developments_2022, guerrero_sp_2009, oogane_subpT_2021, sengupta_temperature_2017}. We assume a typical MTJ p-bit with the following parameters: the free layer magnetization has total magnetization of $M = M_s V$ with volume $V$ and saturation magnetization $M_s$, and the barrier separating the two equilibrium states is $\Delta = KM$ with anisotropy $K$. We shall assume that $\Delta/k_B T \lesssim 10$ for enabling fast thermal fluctuation. In the absence of external magnetic fields, the MTJ p-bit fluctuates between parallel (0) and anti-parallel (1) state with equal probability. 

\begin{figure*}[t]
\centering
\includegraphics[width=\textwidth]{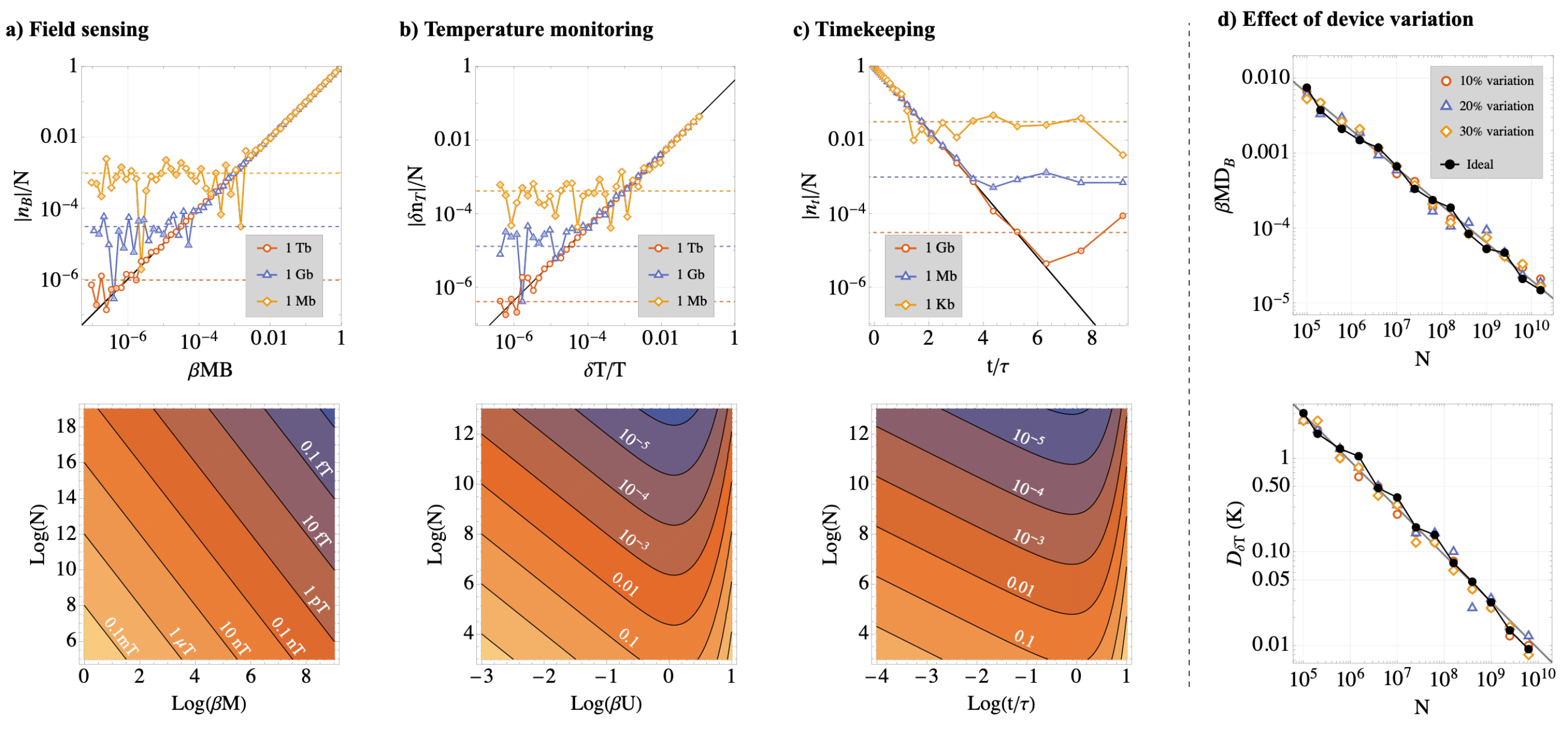}
\caption{a) Upper: Simulation result of imbalance $\abs{n_B}$ as a function of dimensionless field $\beta MB$ for $N = 10^6 (\SI{1}{Mb}), 10^9 (\SI{1}{Gb})$ and $10^{12} (\SI{1}{Tb})$; Lower: Theoretical prediction of field detectivity $D_B$ in \Eq{eqn:DB} as a function of $N$ and total magnetization $\beta M$. The negative $n_B$ values are marked with double open symbols. b) Upper: Simulated change of imbalance $\abs{\delta n_T}$ as a function of relative temperature change $\delta T/T$ for $\beta U=1$ and room temperature ($T=\SI{300}{K}$); Lower: Predicted temperature detectivity $D_{\delta T}/T$ in \Eq{eqn:DT} as a function of $N$ and barrier height $\beta U$ at room temperature. The optimal choice is at $\beta U = 1.2$. c) Upper: Simulated imbalance $\abs{n_t}$ as a function of dimensionless time $t/\tau$; Lower: Predicted temporal detectivity $D_t/t$ in \Eq{eqn:deltat} as a function of $N$ and time $t$. d) The effect of device-to-device variation of p-bits on the overall performance for field sensing (upper) and room temperature monitoring (lower).}
\label{fig:FTt}
\end{figure*}

\emph{Field Sensing -} 
In order to measure weak magnetic fields, the p-bit must respond to the external field. As shown in \Figure{fig:p-measurements}(a), the magnetic field $B$ along the anisotropy direction gives rise to an energy bias $U_B = MB$ between the parallel and anti-parallel state, which leads to an imbalance of the MTJ p-bits
\begin{equation}
    n_B = N \tanh(\beta MB) \simeq N\beta M B
    \equiv\chiB B,
\end{equation}
where $\chiB = N\beta M$ is the susceptibility of the imbalance in response to a small external field $B$.
The statistical fluctuation of the p-bits is roughly 
\begin{equation}
    \delta n 
    = 2\sqrt{N_0N_1/N}
    = \sqrt{N} \sech(\beta MB) \simeq \sqrt{N},
\end{equation}
which equals to $\sqrt{N}$ when the external field is weak.
Therefore, when the field induced imbalance is greater than the fluctuations $\abs{n_B} > \abs{\delta n}$, the external field can be distinguished by the imbalance $n_B$. 
The weakest detectable field, or the field detectivity $D_B$, is given by  
\begin{equation}
    \label{eqn:DB}
    D_B = \frac{\delta n}{\chiB} 
    = \frac{1}{\sqrt{N}}\frac{k_BT}{M_sV}.
\end{equation}
The upper panel of \Figure{fig:FTt}(a) shows the readout of $n_B$ as function of external magnetic field for $N = \SI{1}{Mb}, \SI{1}{Gb}, \SI{1}{Tb}$, respectively. The lower panel of \Figure{fig:FTt}(a) shows the field detectivity \Eq{eqn:DB} as a function of $\beta M$ and $N$.
\Eq{eqn:DB} shows that increasing the total magnetization $M_sV$ is more effective in enhancing the field sensitivity than increasing p-bit number $N$. However, the larger magnetization would suppress the magnetization fluctuation. As shown by Chen \etal \cite{chen_magnetic-tunnel-junction-based_2022}, this contradiction can be resolved by applying a transverse magnetic field to reduce the potential barrier between the 0 and 1 state, thus enhancing the fluctuation rate significantly. Based on a MTJ with $M_s = \SI{9.6e5}{A/m}$ and $V = \SI{e4}{nm^3}$ \cite{Zhao_cofeb_2011, vodenicarevic_cofeb_2017}, $k_BT/M \sim \SI{0.3}{mT}$ in room temperature, thus a field sensitivity of $\SI{1}{nT}$ can be achieved with $N \sim \SI{100}{Gb}$ outputs from p-bits at room temperature.

\emph{Temperature Monitoring -}
We now consider biased p-bits whose parallel and anti-parallel configurations have energy difference of $2U$. The bias can be implemented by some intrinsic pinning within the MTJ or by applying a fixed external field. Because of the energy difference, the imbalance at thermal equiblirium is non-zero (at temperature $T$): 
\begin{equation}
    n_T = N_1(T) - N_0(T) = -N\tanh(\beta U).
\end{equation}
A temperature change $\delta T$ will cause a change in $n_T$ by
\begin{equation}
    \delta n_T = \pdv{n_T}{T}\delta T
    = N\beta U \sech^2(\beta U) \frac{\delta T}{T}
    = \chiT \frac{\delta T}{T}.
\end{equation}
Here $\chiT \simeq N\beta U$ is the susceptibility of the imbalance in response to the temperature variation $\delta T/T$.
In the mean time, the statistical fluctuation in $n_T$ is roughly
\begin{equation}
    \delta n \simeq 2\sqrt{Np_0p_1} 
    = 2\sqrt{N_0N_1/N}
    = \sqrt{N}\sech(\beta U).
\end{equation}
When the temperature-change induced imbalance-change exeeds the statistical fluctuation $\abs{\delta n_T} > \abs{\delta n}$, the temperature change can be inferred from $\delta n_T$. This also defines the minimum temperature change detectable, or the temperature detectivity $D_{\delta T}$
\begin{equation}
    \label{eqn:DT}
    \frac{D_{\delta T}}{T} 
    = \frac{\delta n}{\chi_\ssf{T}}
    = \frac{\cosh(\beta U)}{\beta U \sqrt{N}}
    \gtrsim \frac{1.5}{\sqrt{N}}.
\end{equation}
The lower bound in \Eq{eqn:DT} happens at $\beta U \simeq 1.2$. 
The upper panel of \Figure{fig:FTt}(b) shows the readout of $\delta n_T$ as a function of temperature variation $\delta T$. The lower panel of \Figure{fig:FTt}(b) shows the temperature detectivity \Eq{eqn:DT} as a function of $\beta U$ and $N$, according to which a temperature variation of $\sim$\SI{1}{mK} at room temperature can be detected with $N = \SI{200}{Gb}$ of p-bits and $\beta U=1$.

\emph{Timekeeping -} In order to use p-bits for timekeeping, an ensemble of unbiased p-bits is first initialized by resetting them to state-0 at $t = 0$, such that the initial imbalance $n = N_0 - N_1 = N$ maximizes. Afterwards, the thermalization process will drive the system towards a balanced distribution with $n \sim 0$. During the thermalization process, the imbalance decreases exponentially as function of time:
\begin{equation}
    n_t = N_0(t) - N_1(t) = N e^{-t/\tau},
\end{equation}
where $\tau \propto e^{\beta\Delta}$ is the thermalization time, or the half-life of p-bits. By measuring the imbalance $n_t$ at a later time, the time $t$ elapsed since the initialization can be inferred as:
\begin{equation}
    t = -\tau \ln(n_t/N).
\end{equation}
The statistical fluctuation in $n_t$ is given by 
\begin{equation}
    \delta n_t = 2\sqrt{N_0N_1/N}
    \simeq \sqrt{N(1-e^{-2t/\tau})},
\end{equation}
and the relative uncertainty $\delta t$ (detectivity $D_t$) is 
\begin{equation}
    \label{eqn:deltat}
    \frac{D_t}{t}
    =\frac{\delta t}{t} 
    = \abs{\pdv{t}{n_t}} \frac{\delta n_t}{t}
    = \frac{\tau}{t} \sqrt{\frac{e^{2t/\tau}-1}{N}}, 
\end{equation}
which reaches its minimum when the elapsed time is comparable to half-life of the p-bits. 
The upper panel of \Figure{fig:FTt}(c) shows the readout of $n_t$ as a function of time $t$, which shows that merely $\SI{1}{Kb}$ p-bits can already make a timekeeping on the time scale of one half-life. The lower panel of \Figure{fig:FTt}(c) shows uncertainty in time \Eq{eqn:deltat} as a function of elapsed time $t$ and bit number $N$.

\emph{Efficient read-out -}
Traditionally, the determination of imbalance in a system comprising $N$ p-bits requires reading all p-bits, which is notably time-consuming. Fortunately, since the specific states of the p-bits (whether 0 or 1) are irrelevant to the imbalance, it is feasible to ascertain this information through minimal measurements. By arranging the p-bits in a $P\times Q = N$ series-parallel circuit configuration \cite{guerrero_sp_2009}, the imbalance is intricately linked to the overall resistance of the circuit. 
Let $r$ and $R$ be the resistance of the MTJ p-bit in the parallel (0) and anti-parallel (1) states. Suppose the $i$-th branch has $N_0^i$ p-bits in $r$ and $N_1^i$ in $R$ with $P = N_0^i + N_1^i$ and $n_i \equiv N_1^i - N_0^i \ll P$, and the total resistance of the series-parallel circuit is 
\begin{align}
    R_{sp} 
    = \qty(\sum_{i=1}^Q \frac{1}{N_0^i r + N_1^i R})^{-1}
    \simeq \frac{R+r}{2Q/P}\qty( 1 + \frac{R-r}{R+r} \frac{n}{N}),
\end{align}
which establishes a one-to-one association between $R_{sp}$ and the imbalance $n = \sum_i n_i = N_1 - N_0 \ll N$.
The exact partition of the p-bits for $P$ and $Q$ can be tuned according to the range and precision of resistance measurement. 
This approach not only simplifies the process but also significantly reduces the time and resources needed for determining the imbalance.

\emph{Influence of imperfections of p-bits -}
To achieve higher measurement accuracy, traditional methods typically rely on the enhancement of instrument or device quality. However, the probabilistic measurements discussed in this Letter deviate from this norm. Here, accuracy is improved by increasing the quantity of p-bits rather than by refining the quality of each individual p-bit. This methodology not only proves to be more cost-effective but also enhances flexibility and scalability across various applications. As demonstrated in \Figure{fig:FTt}(d), discrepancies among p-bits, simulated through parameter variations like $M, \beta, B, U$ by up to $\pm 30\%$, do not adversely affect the measurement sensitivity. This robustness stems from the error-cancelling effect achieved when employing a large ensemble of p-bits.

\emph{Discussion -} The concept of probabilistic field sensing hinges on the ability to achieve remarkably high sensitivity levels, contingent upon the acquisition of substantial data volumes, as illustrated in \Figure{fig:FTt}(a). Specifically, with an $N$ value approximating $10^{20}$, or equivalently $\SI{e8}{Tb}$, femto-Tesla sensitivity becomes attainable. This can be realized through repeated measurement of a series-parallel circuit of a p-bit array, where $N$ represents the cumulative total of the p-bits in the circuit multiplied by the number of measurement iterations. Conversely, temperature monitoring demands a significantly lower number of p-bits -- mere gigabits suffice to achieve a sensitivity of $\SI{10}{mK}$ at room temperature. While probabilistic timekeeping may not rival the precision of atomic clocks or conventional timepieces in terms of accuracy, its simplicity in design and robustness against non-uniformity and disturbances make it a viable option. Requiring around $\SI{1}{Kb}$ independent p-bits, probabilistic timekeeping is particularly conducive to biomorphic implementations \cite{buonomano_your_2017}. The discussion thus far has centered on probabilistic measurements utilizing independent p-bits. Given the nature of probabilistic computing, which involves interconnecting p-bits into a network, it prompts an inquiry into how such interconnected p-bits might enhance measurement sensitivity in probabilistic systems.

In conclusion, we proposed an innovative probabilistic measurement framework utilizing an ensemble of p-bits. This approach was demonstrated to be effective in applications such as field sensing, temperature variation monitoring and timekeeping. Notably, by augmenting the total number of p-bits, we can attain unprecedented sensitivity in these domains. Furthermore, the requisite quality and uniformity of the p-bits are not critical factors for achieving high sensitivity, which significantly enhances the scalability of our proposed scheme.

\emph{Acknowledgements -} 
This work was supported by the National Key Research and Development Program of China (Grant no. 2022YFA1403300) and Shanghai Municipal Science and Technology Major Project (Grant No.2019SHZDZX01).

\bibliographystyle{apsrev}
\bibliography{ref}
\end{document}